# Resonant optical trapping of Janus nanoparticles in plasmonic nanoaperture

Alemayehu Nana Koya*, Longnan Li, and Wei Li*

*GPL Photonics Laboratory, State Key Laboratory of Luminescence and Applications, Changchun Institute of Optics, Fine Mechanics and Physics, Chinese Academy of Sciences, Changchun 130033, Jilin, P. R. China*

Authors to whom correspondence should be addressed: alemayehu.koya@gail.com; weili1@ciomp.ac.cn

**ABSTRACT**

Controlled trapping of light absorbing nanoparticles with low-power optical tweezers is crucial for remote manipulation of small objects. This study takes advantage of the synergetic effects of tightly confined local fields of plasmonic nanoaperture, self-induced back-action of nanoparticles, and resonant optical trapping method to demonstrate enhanced manipulation of Janus nanoparticles in metallic nanohole aperture. We theoretically demonstrate that displacement of Au-coated Janus nanoparticles toward plasmonic nanoaperture and proper orientation of the metal coating give rise to enhanced near-field intensity and pronounced optical force. We also explore the effect of resonant optical trapping by employing dual laser system, where an on-resonant green laser excites the metal-coated nanoparticle whereas an off-resonant near-infrared laser plays trapping role. It is found that, at optimum nanoparticle configuration, the resonant optical trapping technique can result in three-fold enhancement of optical force, which is attributed to excitation of surface plasmon resonance in Janus nanoparticles. The findings of this study might pave the way for low power optical manipulation of light-absorbing nanoparticles with possible applications in nanorobotics and drug delivery.

Janus nanoparticles (JNPs), small composite objects having two or more parts with distinct chemical and physical properties,[1] have been widely explored for novel applications in biosensing,[2] cancer diagnosis,[3] and nanoscale robotics.[4] The compositional asymmetry of Janus nanoparticles gives unprecedented opportunities to manipulate these architectures with optical manipulation techniques, which are attractive because of their fuel-free and remote operation. As a result, several works have been reported on optical manipulation of Janus nanoparticles with, for example, focused optical tweezers[5] and free-space beams.[6] However, optical trapping of metallic nanoparticles and metal-coated Janus nanoparticles are very challenging, as the high reflectance and absorbance of metals lead to strong forces that repel the particles from the optical trap.[7, 8] Nevertheless, since metallo-dielectric particles in optical fields tend to move toward higher-gradient areas,[9, 10] JNPs can be easily manipulated with near-field optical trapping techniques. Due to very tight confinement of light and associated local field enhancement, near-field based optical manipulation schemes have shown promising results in manipulation of light absorbing nanoparticles.[11, 12]

In particular, plasmonic optical tweezers (POTs) take advantage of the near-field effects of metallic nanostructures to manipulate nano-sized objects with low-power.[13, 14] Unlike the diffraction-limited conventional optical tweezers,[15-20] plasmonic optical tweezers rely on the potential of metallic nanostructures to confine light into deep sub-wavelength spaces,[21-23] which makes POTs indispensable tools for dynamic manipulation of various objects, including nanoparticles, cells, and DNA molecules.[24-31] In particular, among several architectures of plasmonic nanotweezers,[26,32] sub-wavelength nanoapertures engraved in thin metallic films[33] have been widely exploited for single molecule studies,[34,35] biosensing,[36] and contactless manipulation of nanometer-sized particles with low power.[11, 37-39]

On the other hand, the efficiency of optical manipulation is also affected by parameters related to nanoparticles, including nanoparticle position and its composition (hence polarizability).[40,41] In SIBA trapping, the particle motion couples to the resonance frequency of the nanoaperture, which results in a strong interplay between the nanoaperture field intensity and the forces exerted.[42] Furthermore, the trapping laser parameters (such as laser intensity and wavelength) also affect trapping of metallic and metal coated nanoparticles.[43] As a general rule, to mitigate the photothermal heating effect of metal nanoparticles, an off-resonant laser is often employed to trap light-absorbing nanoparticles.[8] However, this approach limits the interaction of light with the nanoparticles and it also disregards the SIBA effect. To overcome this limitation, dual laser system is employed, where one laser excites the nanoparticle whereas the other laser plays trapping role.[44-46] Thus, the resonant optical trapping method has been employed to efficiently manipulate both metallic[44,45] and non-metallic particles.[46] In particular, plasmon resonance-based optical trapping method is used to achieve stable trapping of metallic nanoparticles, as surface plasmon modes of these particles enhance the gradient force of an optical trap, thereby increasing the strength of the trap potential.[47]

Nevertheless, the potential of plasmonic nanoapertures for manipulation of light absorbing nanoparticles has not been exploited. In fact, there is a theoretical report on optical trapping of metallic particles with plasmonic nano-apertures.[48] However, this work has disregarded the effect of surface plasmon resonance of trapped metallic nanoparticles. Other related works on plasmonic nanoaperture tweezers have put much effort on designing and optimization of the optical responses of the nanoapertures for enhanced optical trapping applications.[49,50] As a result of these fundamental studies, the underlying physics of nanoaperture-based plasmonic nanotweezers is well understood and the current research focus has thus shifted toward exploiting their potentials for various applications.[51,52] Regardless of these developments, the potentials of plasmonic nanoapertures for optical manipulation of Janus nanoparticles has not been exploited, which would be interesting as the compositional asymmetry and distinct physical properties of the JNPs give additional degree of freedom for low-power optical manipulation. Moreover, detailed study on the self-induced back-action (SIBA) effect of metal-coated nanoparticles trapped in plasmonic nanoapertures might lay foundations for in-depth understanding of optical manipulation of light absorbing nanoparticles.[53]

To this end, we report on resonant optical trapping of light absorbing nanoparticles in a plasmonic nanoaperture, where the plasmonic near-field effect on resonant optical trapping of Au-coated Janus nanoparticles in single nanohole (SNH) Au aperture is studied. In this regard, by moving the Janus nanoparticles toward the SNH aperture, we have investigated self-induced back-action effect of nanoparticles on the near-field dynamics and optical force of the studied nanosystem. In addition, by taking advantage of the compositional asymmetry of JNPs, we have also explored the effects of nanoshell orientation and thickness on optical trapping of such nanoparticles in a SNH plasmonic nanoaperture. Furthermore, using two-laser system, we have also studied the impact of resonant excitation on optical trapping of Janus nanoparticles in plasmonic nanoaperture. The findings of this study might have implications for efficient manipulation of asymmetric nanoparticles with possible applications in nanorobotics and drug delivery.

For our theoretical study, we designed a plasmonic optical tweezer made of circular shaped single nanohole aperture (with a diameter of 300 nm) engraved in 100 nm thick Au film placed on top of 500 nm thick silica substrate (see Fig. 1(a) & (b)). It has been demonstrated that plasmonic optical tweezers are capable of stable trapping nano-sized objects[14,54-56] so that the nano-aperture design parameters were adopted from recent works on single plasmonic nanoaperture tweezers.[37,57] Moreover, to mimic biologically relevant environment for our simulation, we set water (n = 1.33) as a background medium. To investigate the self-induced back-action effect, we used spherical silica nanoparticles (SNPs) and Janus nanoparticles composed of 90 nm silica core and 5 nm thick half-coated Au cap (see Fig. 1(c)). To accommodate the nanoparticle inside the Au SNH aperture, we kept the diameter of the studied nanoparticles within the height of the nanoaperture (see Table S1 in the Supplementary Material).

To induce enhanced plasmonic effect in the SNH Au aperture,[58] the whole system was illuminated with a linearly polarized plane wave centered at 1064 nm. Additionally, to investigate the effect of resonant excitation on the optical trapping, we also used a 532 nm laser that was simultaneously illuminated along the z axis (see Fig. 1(a)). Initially, both lasers had a default optical power of 2.652 mW and the two lasers were uniformly illuminated from the glass side. In the dual-laser system, the green laser excites the nanoparticles whereas the near-infrared (NIR) laser ensnares the nanoparticles. To calculate the near-field dynamics and optical forces of the studied nanosystem, we employed the finite-difference time domain (FDTD) method with perfectly matched layer boundary conditions. During the simulation, the refractive index of silica was set to be 1.45 and the dielectric permittivity for the gold aperture and nanoshell was taken from ref.[59] The total optical force acting on the nanoparticles was obtained by integrating the Maxwell

Stress Tensor (MST) around a boundary enclosing the nanoparticles.[6,12,60] For further details on the simulation methods of near-field intensity and optical force, see the Supplementary Material.

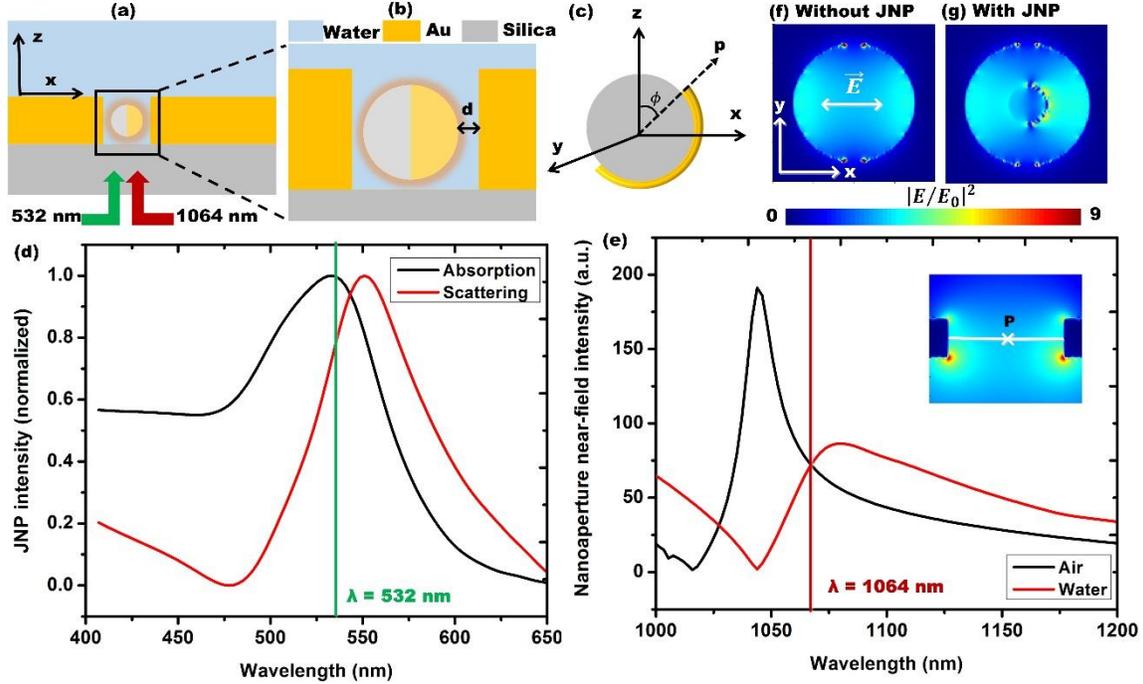

**FIG. 1**. (a) Design principle of near-field mediated resonant optical trapping of Janus nanoparticles (JNPs) in a single nanohole (SNH) Au aperture. (b) Zoomed schematic illustration of the studied nanosystem shown in (a). (c) Schematic illustration of the studied Janus nanoparticle made of silica core and half-coated Au nanoshell, where the metallic coating is rotated by an angle $\phi$. (d) Absorption and scattering resonances of the half-Au coated (5 nm thick) silica core (90 nm) Janus nanoparticle. (e) Near-field intensity of the SNH Au aperture calculated at its geometrical center for the displayed background media. The inset shows near-field profile of the SNH aperture with water background. (f) and (g) denote near-field profiles of the SNH aperture calculated in absence and presence of JNP, respectively.

With the aforementioned theoretical bases and design principles, we have studied self-induced back-action effect of resonantly excited Janus nanoparticles by changing their position and orientation within the single nanohole plasmonic aperture. To explore these effects, first, we calculated the optical responses of the studied nanosystem (Fig. 1(d) – (g)). Silica-core Au-half coated Janus nanoparticle with a maximum diameter of 100 nm has an absorption resonance around 532 nm (Fig. 1(d)), which is resonant with the green laser used to excite the nanoparticle. In absence of the JNP, the SNH aperture shows enhanced near-field intensity in air than in water (Fig. 1(e)). However, introducing JNP into the cavity region of the nanoaperture gives rise to pronounced near-field intensity (compare Fig. 2 (f) & (g)), which is attributed to the self-induced back action effect.[61] It is a well-established fact that introducing a small object into the hotspot region of sub-wavelength plasmonic nanocavities results in redshift of the resonance peak of the nanosystem[43, 62] and significant enhancement in the near-field intensity.[37] With the SIBA effect, it has been able to trap sub-100 nm particles with low power.[11, 42] Thus, our SNH aperture tweezer benefits from the SIBA effect of the JNP where presence of trapped nano-object helps to further improve the restoring force.

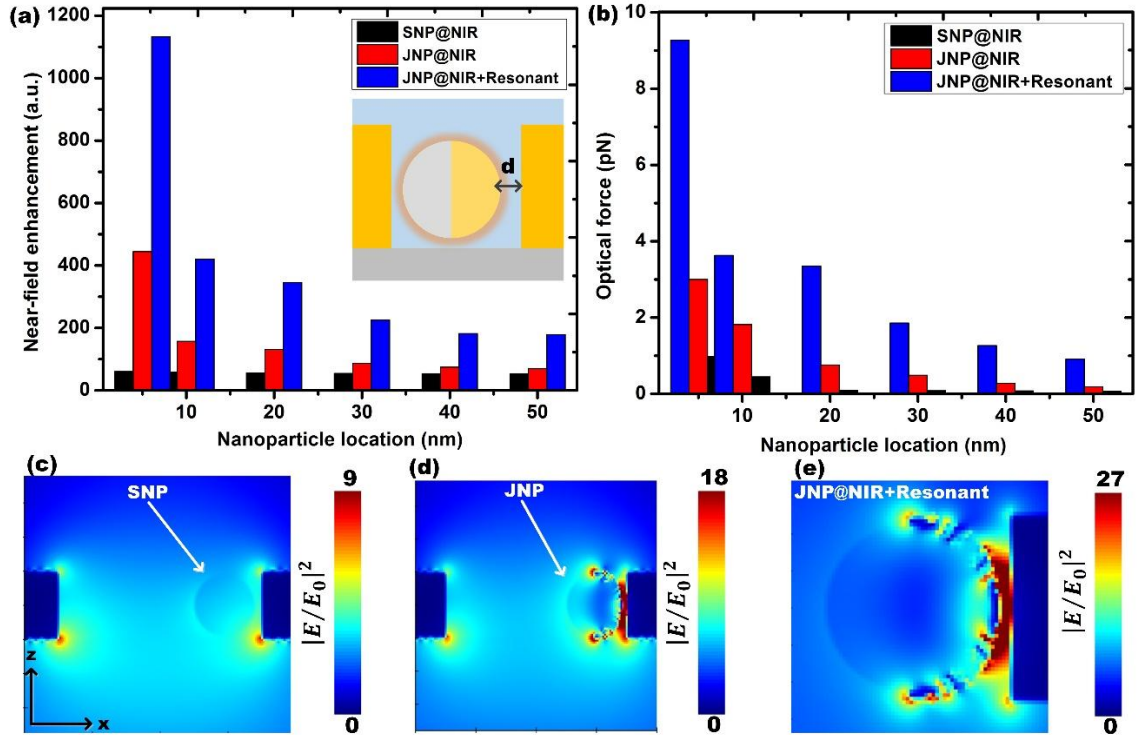

**FIG. 2**. (a) The effect of nanoparticle location on the near-field intensity of SNH aperture achieved with single laser (NIR) and dual-laser system (NIR + Resonant) for optical trapping of silica nanoparticle (SNP) and Janus nanoparticles. The inset schematic illustrates the location of studied nanoparticles with respect to the SNH aperture. (b) Corresponding optical forces acting on the SNP and JNP. (c) and (d) show SNH aperture near-field profiles in presence of SNP and JNP located at 5 nm, respectively. (e) Zoomed near-field profile of JNP attained with the two-laser system calculated for the nanoparticle located at 5 nm away from the SNH aperture.

However, it was found that when the Janus nanoparticle is placed at the geometric center of the SNH aperture, which is 100 nm away from the SNH aperture and approximately equal to the nanoparticle size, the presence of the particle has a little effect on the near-field intensity of the nanoaperture. Thus, to take full advantage of the SIBA effect, the JNP was moved toward the SNH aperture along the x-axis and the corresponding near-field intensity of the studied nanosystem calculated at 1 nm away from the SNH aperture is displayed in Fig. 2(a). As result of the presence of 5 nm thick Au nanoshell, the near-field intensity of the nanocavity in presence of the JNP is found to be about seven times as large as that achieved in presence of bare silica nanoparticle having comparable size (see black and red bars in Fig. 2(a)). The impact of Au coating is also clearly visible in the optical forces calculated for both SNP and JNP, where the latter composition gives rise to more than two times enhanced optical force strength compared to the former nanoparticle (compare black and red bars in Fig. 2(b)). These results imply that the SIBA effect in optical trapping of nano-sized objects can be further enhanced by coating nanoparticles with thin metallic layers.

On the other hand, recent works have demonstrated that the optical resonance between a light-absorbing particle and trapping laser can give rise to about four-fold enhancement of radiation force, which is attributed to the excitation of surface plasmon resonance.[45,46] Thus, we have also investigated the effect of resonant excitation of Janus nanoparticles on near-field dynamics and optical forces by employing a dual-laser system, where 532 nm laser resonantly excites localized surface plasmon in the JNP and 1064 nm laser traps the nanoparticle. Under this condition, we calculated both the near-field intensity and optical force of the JNP-in-nanoaperture system (see blue bars in Fig. 2(a) & (b)). The nanoaperture near-field intensity calculated for resonant optical trapping is found to be more than two times larger than that achieved with single NIR laser. Moreover, using two laser beams where one laser resonant with the JNP absorption gives rise to about three-fold enhancement of optical force. This is due to the fact that, when the photon energy of the laser field matches the excitation bands of the materials, the radiation force can be drastically altered through resonantly enhanced induced polarization.[63] This effect is more prominent especially when trapping light-absorbing materials such as metallic nanoparticles and metal-coated Janus nanoparticles (see Fig. 2(c) – (e)). Thus, the challenges associated with stable trapping of light absorbing nanoparticles can be overcome by employing two-laser system, one off-resonant tightly focused trapping laser and another on-resonant unfocused excitation laser beam.[45, 46]

Furthermore, recent works on optical manipulation of Janus nanoparticles have demonstrated that the orientation of metal coating with respect to trapping beam affects the strength of the optical forces acting on the nanoparticles.[6,12] By extending these efforts for plasmonic trapping of light absorbing nanoparticles, we have investigated the impact of Au nanoshell orientation on the near-field dynamics and optical forces of the JNP-in-nanoaperture system. As illustrated in Fig. 3(a), the orientation of half-coated Au nanoshell was changed with respect to the wall of the nanoaperture. When the Au coating is placed at the vicinity of the nanoaperture, the local field intensity of the studied nanosystem shows significant increment (see Fig. 3(b) & (c)). Expectedly, the optical force of the nanosystem as a function of the Au coating orientation follows the same trend as the near-field dynamics. In particular, those configurations with closest orientation of the Au nanocap and its edge to the nanoaperture ($\phi = 0°, 90°, 270°$ and $360°$) provide better contribution to the enhancement of the nanocavity near-field intensity and optical force. Among them, the orientation of the nanoshell at $90°$ gives the highest near-field enhancement and optical force strength. This can be attributed to the full exposure of the metallic coating to the impinging light and at the same time closest orientation of the Au nanocap to the aperture. On the other hand, for the nanoshell configurations where the metallic coating is oriented away from the nanoaperture (see $\phi = 135°, 180°$, and $225°$), the optical force acting on the nanoparticle is found to be the least one. This implies that stable trapping of Janus nanoparticles in plasmonic nanoapertures can be realized for orientations of the metallic coating far away from the nanoaperture, where the plasmonic effect of the nanocap is minimum.

Overall, metallic coating of nanoparticles with nearer orientation to the SNH aperture results in enhanced cavity near-field intensity and optical force, which further confirms that the enhanced near-field intensity and optical force displayed in Fig. 2 (a) & (b) and Fig. 3(b) stem from the resonant excitation of surface plasmons in the Janus particle but not because of the relatively higher optical power coming from two laser beams. In fact, this result is consistent with recent study reported on optical manipulation of Janus particles along optical nanofibers, where the optical force acting on the particle found to be maximum for the gold cap orientations closer to the evanescent field of the optical nanofiber.[12] Furthermore, related work on plasmonic tip based optical trapping of a polystyrene nanoparticle with gold cap reported that the metal tip can capture the particle at the position of the gold cap due to the strong plasmonic interaction, while other positions of the particle cannot be captured by metal tip.[64]

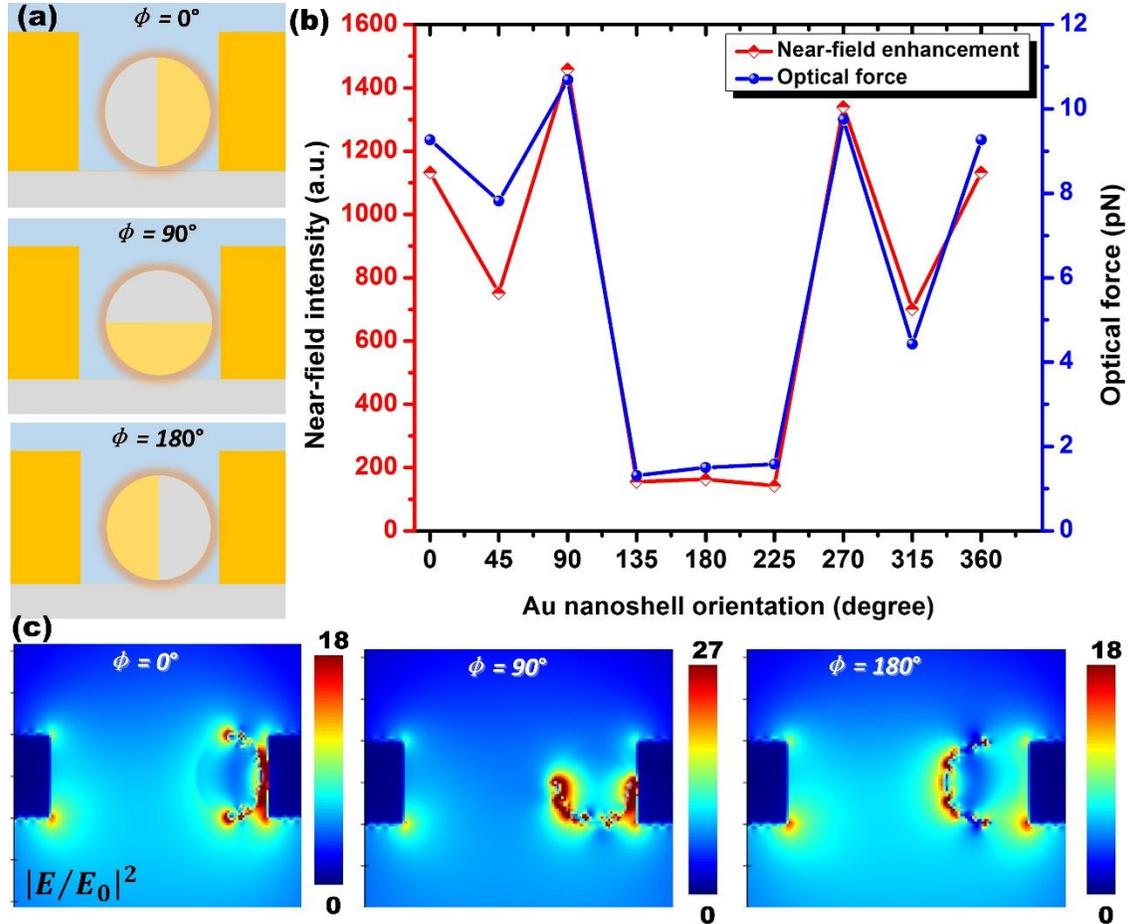

**FIG. 3**. The effect of Au nanoshell orientation on the near-field dynamics and optical force of SNH Au aperture containing Janus nanoparticle. (a) Schematic illustration of various orientations of the Au nanocap inside the nanoaperture. (b) Variations of the near-field intensity and optical force as a function of the JNP Au nanoshell orientation. (c) Near-field profiles of the JNP-in-nanoaperture system for different Au cap orientations.

Finally, we have explored the effects of Au nanoshell thickness and optical power on the near-field intensity and optical force of the studied nanosystem. As displayed in Fig. 4(a), both the near-field intensity and optical force of the nanosystem show maximum enhancement as the nanoparticle composition changes from bare silica nanoparticle to Au coated Janus nanoparticle with shell thickness of 10 nm. However, as the nanoshell thickness further increases, the near-field intensity and optical force tend to decline gradually and show tendency of saturation. Indeed, in related work reported on optical manipulation of Janus particles with the evanescent field of optical fiber,[12] the optical force acting on Janus particles attains its maximum value as the Au coating thickness approaches 30 nm, which is close to the skin depth of gold.[65] Similarly, it has been theoretically demonstrated that the optical force acting on a single gold nanoparticle in human serum albumin solutions decreases with increasing the nanoparticle volume fraction.[66] These findings imply that near-field based manipulation of Janus particles made of dielectric core and metal shell benefits from both the trapping stability and thermal neutrality of the dielectric core and the high polarizability of the metallic coating.[12]

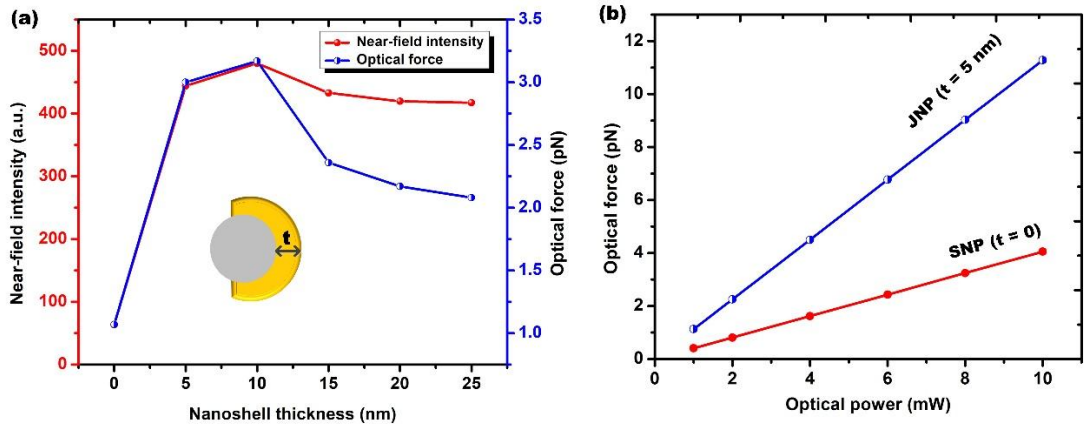

**FIG. 4**. (a) The effect of gold nanoshell thickness $t$ on the near-field intensity and optical force of the studied nanosystem illuminated by 2.652 mW NIR laser. (b) Scaling of the trapping NIR laser power with the optical force exerted on the Janus nanoparticle ($t = 5$ nm) and silica nanoparticle ($t = 0$) trapped in plasmonic nanoaperture.

Moreover, as shown in Fig. 4(b), for both SNP and JNP, the optical force strength increases linearly as the trapping laser power increases from 1 mW to 10 mW. In fact, this linear relationship with optical power was also observed for propulsion speed (and hence optical force) of Janus particles trapped with the evanescent field of optical nanofiber.[12] However, increasing the optical power of laser beam to attain higher radiation pressure might damage nanoparticles, especially those with lower damage thresholds. In particular, when trapping biological samples by means of plasmonic tweezers, the net optical power should not exceed the damage threshold of the specimens.[67] Alternatively, as theoretically shown in this study, one can attain comparable optical force with minimum input power by exploiting the rotational degree of freedom of Janus particles, where presence of metallic coating gives rise to enhanced optical force (see Table I). However, in the real-experiment configurations where randomly positioned nanoparticles are trapped with tightly focused beam in fluidic environment, in addition to the light-induced radiation pressure, one should also take into account other dominant effects such as thermophoresis, capillary action, and hydrodynamic drug force.[5, 68]

**TABLE I.** Comparison of the near-field enhancement (NF) and optical force (F) of Janus nanoparticles trapped in plasmonic optical tweezers under various trapping conditions

| Trapping conditions | | | | Optimum results | | Reference figure |
|---|---|---|---|---|---|---|
| **Parameters** | **Laser system** | **λ (*nm*)** | **P (*mW*)** | **NF [*a.u.*]** | **F [*pN*]** | |
| Location effect | Double | 532, 1064 | 1.628 | 1132.65 | 9.27 | 2(a), (b) |
| Rotation effect | Double | 532, 1064 | 1.628 | 1458.94 | 10.69 | 3(b) |
| Thickness effect | Single | 1064 | 1.628 | 444.20 | 3.00 | 4(a) |
| Optical power effect | Single | 1064 | 10 | 861.64 | 11.28 | 4(b) |

In this study, we have investigated resonant optical trapping of Janus nanoparticles in a gold nanoaperture as a plasmonic optical tweezer. In this regard, we have explored the effect of self-induced back-action of Janus nanoparticles on the near-field dynamics and optical forces of the studied nanosystem. In addition, we have also investigated the impacts of nanoparticle rotation and composition on the optical trapping of Janus nanoparticles. We have theoretically demonstrated that, with resonant optical trapping method and optimum nanoparticle configuration, one can achieve superior optical trapping efficiency for Janus nanoparticles in plasmonic nanoapertures. These findings imply that, to take full advantage of the plasmonic effects of metallic nanoparticles for novel applications,[69-72] it is important to design functional nanostructures with metallic components so that one can optically manipulate[73,74] such a composite nanomaterials in fluidic environments.[75,76] Moreover, the controlled and low-power manipulation of Janus nanoparticles shown in this study is expected to find applications in nanorobotics and drug delivery.

**SUPPLEMENTARY MATERIAL**

See the supplementary material for the detailed description of simulation methods of the near-field intensity and optical force.

This work was supported by National Natural Science Foundation of China (Grant No. 62134009 and 62121005) and Chinese Academy of Sciences President's International Fellowship Initiative (Grant No. 2023VMC0020).

**AUTHOR DECLARATIONS**

**Conflict of Interest**

The authors have no conflicts to disclose.

**DATA AVAILABILITY**

The data that support the findings of this study are available from the corresponding authors upon reasonable request.

# Supplementary Material

# Resonant optical trapping of Janus nanoparticles in plasmonic nanoaperture

Alemayehu Nana Koya*, Longnan Li, and Wei Li*

*GPL Photonics Laboratory, State Key Laboratory of Luminescence and Applications, Changchun Institute of Optics, Fine Mechanics and Physics, Chinese Academy of Sciences, Changchun 130033, Jilin, P. R. China*

Authors to whom correspondence should be addressed: alemayehu.koya@gmail.com; weili1@ciomp.ac.cn

**Simulation Methods for near-field intensity and optical force**

To investigate the resonant optical trapping of light absorbing nanoparticles in plasmonic nanoapertures, we used spherical Janus nanoparticles (JNPs) composed of 90 nm silica core and 5 nm thick half-coated Au cap. Moreover, to compare the optical properties of the studied nanoparticles, we have also used silica nanoparticles (SNPs) of comparable size. Nevertheless, to accommodate the nanoparticles inside the 100 nm thick Au SNH aperture, we kept the diameter of the studied nanoparticles within the height of the nanoaperture. The studied nanosystems have the following parameters.

**Table S1**. Geometrical parameters of studied nanosystems.

| Nanosystem | Shape | Composition | Diameter (nm) | Thickness (nm) |
|---|---|---|---|---|
| Plasmonic nanoaperture | Circular | Au | 300 | 100 |
| Janus nanoparticle | Spherical | Silica-core, Au-nanoshell | 100 | 100 |
| Silica nanoparticle | Spherical | silica | 100 | 100 |

The near-field responses and optical forces of the studied nanosystem were simulated with the finite-difference time domain (FDTD) method with perfectly matched layer boundary conditions. The refractive index of silica was set to be 1.45 and the dielectric permittivity for the gold aperture and nanoshell was chosen from ref.[1] To ensure the accuracy of the results, we used mesh size of 2 nm in all three directions uniformly. The near-field intensities of the bare nanoaperture and the nanoparticle-in-nanoaperture system, respectively, were calculated at the geometric center of the nanoaperture (x = y = z = 0) and at 1 nm (x = 149 nm, y = z = 0) away from the circular nanoaperture. The optical forces acting on the nanoparticles were obtained by integrating the Maxwell Stress Tensor around a boundary enclosing the nanoparticle. To ensure its accuracy, we carefully set the MST geometry only to enclose the studied nanoparticles and not cross any other surfaces.

The total optical force acting on the nanoparticles was obtained by integrating the Maxwell Stress Tensor (MST) around a boundary enclosing the nanoparticle, which is given by[2]

$$\langle \mathbf{F} \rangle = \oint \langle \mathbf{T} \rangle \cdot \hat{\mathbf{n}} \, dA \tag{S1}$$

where $\langle \ \rangle$ denotes the time average, $\hat{\mathbf{n}}$ is a unit vector normal to the enclosing surface $dA$, the surface enclosing the particle and $\mathbf{T}$ is the MST given by[3, 4]

$$T_{ij} = \varepsilon_m \left( E_i E_j - \frac{1}{2} |\mathbf{E}|^2 \delta_{ij} \right) + \frac{1}{\mu_0} \left( B_i B_j - \frac{1}{2} |\mathbf{B}|^2 \delta_{ij} \right) \tag{S2}$$

here $\varepsilon_m$ is the permittivity of surrounding medium, $\mathbf{E}$ and $\mathbf{B}$, respectively, denote electric and magnetic fields, $\mu_0$ is the vacuum permeability, and $\delta_{ij}$ is the Kronecker delta.